\documentclass[prl,aps,showpacs,superscriptaddress,twocolumn]{revtex4-1}
\usepackage{latexsym}
\usepackage{amsfonts}
\usepackage{amssymb}
\usepackage{amsmath}
\usepackage{graphicx}
\begin{document}
\bibliographystyle{apsrev}
\title{Measurement of a Sign-Changing Two-Gap Superconducting Phase in Electron-Doped\\
 $\rm Ba(Fe_{1-x}Co_x)_2As_2$ Single Crystals using Scanning Tunneling Spectroscopy}
\date{\today}

\author{M.~L. TEAGUE$^1$,  G. K. DRAYNA$^1$, G. P. LOCKHART$^1$, P. CHENG$^2$, B. SHEN$^2$, H.-H. WEN$^2$ and N.-C. YEH}

\affiliation{Department of Physics, California Institute of Technology, Pasadena, CA 91125 \\
$^2$Institute of Physics, Chinese Academy of Sciences, China}

\begin{abstract}
Scanning tunneling spectroscopic studies of $\rm Ba(Fe_{1-x}Co_x)_2As_2$ (x = 0.06, 0.12) single crystals reveal direct evidence for predominantly two-gap superconductivity. These gaps decrease with increasing temperature and vanish above the superconducting transition $T_c$. The two-gap nature and the slightly doping- and energy-dependent quasiparticle scattering interferences near the wave-vectors $(\pm \pi , 0)$ and $(0, \pm \pi)$ are consistent with sign-changing $s$-wave superconductivity. The excess zero-bias conductance and the large gap-to-$T_c$ ratios suggest dominant unitary impurity scattering. 
\end{abstract}
 \maketitle

The recent discovery of iron-based superconductors~\cite{Kamihara08,Takahashi08,ChenXH08,WenHH08,RenZA08,Sasmal08,Rotter08,Sefat08,MasseeF09,KatoT09} has renewed intense research activities in superconductivity. Comparison of the similarities and contrasts between the iron-based compounds and the cuprates can provide useful insights into the microscopic mechanism for high-temperature superconductivity~\cite{Cvetkovic09}. In particular, antiferromagnetic spin fluctuations appear to influence the physical properties of both cuprate and iron-based superconductors~\cite{Cvetkovic09,delaCruz08,SinghDJ08,Haule08,WangF09a,WangF09b}. On the other hand, theoretical calculations~\cite{Sefat08,WangF09a,WangF09b,MazinII08} and angle resolved photoemission spectroscopy (ARPES)~\cite{LuDH08,DingH08,TerashimaK09} suggest that multi-bands and inter-Fermi surface interactions are crucial to superconductivity in the iron-based compounds. 

While substantial experimental results from ARPES studies~\cite{LuDH08,DingH08,TerashimaK09} and phase sensitive measurements of various single crystalline iron-based compounds~\cite{ChenCT10,HanaguriT10} are supportive of the scenario of two-gap superconductivity with sign-changing $s$-wave ($s^{\pm}$) order parameters for the hole and electron Fermi pockets, tunneling and point-contact spectroscopic studies of the iron pnictides appear to be inconclusive~\cite{ChenTY08,ShanL08,BoyerMC08,YinY09,ChuangTM10,FasanoY10}. For instance, reports of point-contact spectroscopy and scanning tunneling spectroscopy (STS) have suggested either BCS-like $s$-wave superconductivity~\cite{ChenTY08} or nodes in the superconducting order parameter~\cite{ShanL08,FasanoY10} of the ``1111'' iron-arsenides $\rm LnFeAsO_{1-x}F_x$ (Ln: trivalent rare-earth elements). On the other hand, STS studies of the electron and hole-doped ``122'' iron arsenides $\rm Ba(Fe_{1-x}Co_x)_2As_2$ and $\rm (Ba_{1-x}K_x)Fe_2As_2$ have revealed findings ranging from strong spatial variations in the tunneling spectra with occasional observation of a large superconducting gap~\cite{BoyerMC08} to moderate spatial variations with predominantly a small superconducting gap~\cite{YinY09}. In the non-superconducting limit, nematic surface reconstructions have been observed~\cite{ChuangTM10}. 

In this letter we report direct STS evidence for two-gap superconductivity in the electron-doped 122 system $\rm Ba(Fe_{1-x}Co_x)_2As_2$ of two different doping levels. For each doping level, two different energy gaps can be clearly resolved at $T \ll T_c$. Both gaps decrease monotonically with increasing temperature and then completely vanish above $T_c$. The gap values agree favorably with those obtained from ARPES so that the larger gap $\Delta _{\Gamma}$ may be associated with the hole-like Fermi pockets and the smaller gap $\Delta _{\rm M}$ with the electron-like Fermi pockets. Moreover, Fourier transformation (FT) of the tunneling conductance reveals energy and doping dependent quasiparticle scattering interferences (QPI) near the nesting wave-vectors $(\pm \pi , 0)$ and $(0, \pm \pi)$ between the Fermi pockets at $\Gamma$ and M, which is consistent with the sign-changing order parameters for the hole and electron pockets~\cite{MazinII08}. Finally, excess zero-bias tunneling conductance and the large $2\Delta _{\Gamma , \rm M}/(k_B T_c)$ ratios for both doping levels may be attributed to strong unitary impurity scattering~\cite{BangY09}.

The $\rm Ba(Fe_{1-x}Co_x)_2As_2$ samples investigated in this work are single crystals with x = 0.06 (underdoped) and 0.12 (overdoped), and the corresponding superconducting transition temperatures are $T_c =$ 14 and 20 K, respectively. The single crystals were grown from the flux method~\cite{FangL09}, and details of the synthesis and characterization of the samples have been described elsewhere~\cite{FangL09,BernhardC09,ShenB10}. Given the reactive nature of $\rm Ba(Fe_{1-x}Co_x)_2As_2$, freshly cleaved surfaces were essential for the STS studies. To date all reported STS studies were carried out on samples that were mechanically cleaved under ultra-high vacuum (UHV) conditions and at cryogenic tempertures~\cite{ChuangTM10,BoyerMC08,YinY09}, and the resulting sample surfaces all exhibited significant $(2 \times 1)$ reconstructions of the Fe(Co)-layer. We chose to perform mechanical cleavage of the single crystals in pure argon atmosphere at room temperature, well above the tetragonal-to-orthorhombic structural phase transition. The cleaved samples were loaded {\it in situ} onto the cryogenic probe of our homemade scanning tunneling microscope (STM) in argon. The sealed STM assembly was subsequently evacuated and cooled to 6 K in UHV with a base pressure at $\sim 10^{-10}$ Torr.   

Both spatially resolved topography and tunnelling conductance ($dI/dV$) versus energy ($E = eV$) spectroscopy were acquired pixel-by-pixel over an extended area of each sample simultaneously, with tunneling currents along the crystalline c-axis. The typical junction resistance was kept at $\sim 1$ G$\Omega$. To remove slight variations in the tunnel junction resistance from pixel to pixel, the differential conductance at each pixel is normalized to the polynomial fit to its high-energy conductance background from $|E| = \Delta _{\rm max} + 1.5$ meV to $|E| = \Delta _{\rm max} + 6.5$ meV. Detailed survey of the surface topography and tunneling conductance spectra was carried out over typically $\rm (5.4 \times 5.4) \ nm^2$ and $\rm (6.0 \times 6.0) \ nm^2$ areas, and each area was subdivided into $(128 \times 128)$ pixels. Generally the tunneling spectra appeared to be relatively consistent throughout each scanned area, with representative point spectra shown in the left panels of Figs. 1(a) and 1(b) for doping levels x = 0.06 and 0.12, respectively. Two predominant tunneling gap features are apparent for both doping levels. In comparison with the ARPES data~\cite{LuDH08,DingH08,TerashimaK09}, we may assign the larger gap to the superconducting gap of the hole Fermi surface at the $\Gamma$-point of the Brillouin zone, $\Delta _{\Gamma}$, and the smaller gap to that of the electron Fermi surface at the M-point, $\Delta _{\rm M}$. However, upon closer inspection, we note that the larger gap features often exhibit broadening or even slight splitting, as exemplified in Figs.~1(a)-(b) and 3(c). The physical origin of this splitting/broadening is unknown.   

\begin{figure}
\centering
\includegraphics[width=3.4in]{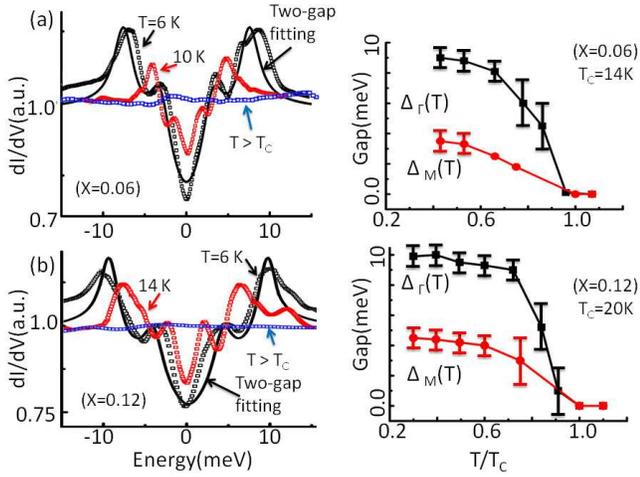}
\caption{(color online) Direct spectroscopic evidences for two-gap superconductivity in $\rm Ba(Fe_{1-x}Co_x)_2As_2$: {\bf (a)} Left panel: normalized tunneling conductance ($dI/dV$) vs. bias voltage ($V$) spectra taken at $T$ = 6, 8 and 15 K for the sample with x = 0.06 and $T_c = 14$ K. The solid lines represent theoretical fittings to spectra using the Dynes formula in Eq.~(1) modified for two-gap BCS superconductors. Two distinct tunneling gaps $\Delta _{\Gamma}$ and $\Delta _{\rm M}$ can be identified from the spectrum at $T$ = 6 K. Right panel: The tunneling gaps $\Delta _{\Gamma}$ and $\Delta _{\rm M}$ as a function of the reduced temperature $(T/T_c)$ are shown by the symbols and solid lines. The error bars indicate the widths of the gap distributions obtained from the fitting using Eq.~(1). {\bf (b)} Left panel: ($dI/dV$) vs. ($V$) spectra taken at $T$ = 6, 10 and 21 K for the sample with x = 0.12 and $T_c = 20$ K. Right panel: $\Delta _{\Gamma , \rm M}$-vs.-$(T/T_c)$.}
\label{fig1}
\end{figure}

Next, we employ a phenomenological fitting generalized from the Dynes formula~\cite{DynesRC78} to analyze the spectra and we restrict to two-gap superconductivity in our analysis. Specifically, the normalized tunneling conductance $\bar{G}$ for a metal-insulator-superconductor junction in the case of a two-gap superconductor may be given by: 
\begin{equation}
\bar{G} = A + \sum_ {i = \Gamma , \rm M} B_i \int {\rm Re} \left[ \frac{(E - i \Gamma _i)(df/dE)|_{E-eV}}{\sqrt{(E - i \Gamma _i)^2 - \Delta _i ^2}}\right] dE.
\label{eq:Gfit}
\end{equation}
Here $A$ and $B_i$ are positive constants, $\Gamma _i$ denotes the quasiparticle scattering rate associated with the superconducting gap $\Delta _i$, and $f(E)$ is the Fermi function. Hence, by applying Eq.~(1) to the temperature dependent tunneling conductance in the left panels of Figs.~1(a)-(b), we obtain temperature dependent values for $\Delta _{\Gamma}$ and $\Delta _{\rm M}$, which are illustrated in the right panels of Figs.~1(a)-(b). Both gaps are particle-hole symmetric (see Fig.~2) and vanish immediately above $T_c$ for both doping levels, implying that $\Delta _{\Gamma}$ and $\Delta _{\rm M}$ are indeed superconducting gaps. Further, the quasiparticle scattering rates derived from Eq.~(1) are very large even at $T = 6$ K, showing $(\Gamma _{\Gamma}/\Delta _{\Gamma}) = 0.4$ and 0.5 for x = 0.06 and 0.12, and $(\Gamma _{\rm M}/\Delta _{\rm M}) = 0.1$ for both x = 0.06 and 0.12. 

It is apparent from Fig.~1 that the two-gap fitting is not ideal, which may be attributed to the following. First, the generalized Dynes formula does not explicitly consider the possibility of a sign-changing $s$-wave order parameter, the latter is theoretically shown to be very sensitive to unitary impurities so that the zero-bias conductance may be strongly enhanced without requiring a large $(\Gamma/\Delta)$ ratio~\cite{BangY09}. Second, there may be different gaps associated with the two hole-pockets, so that Eq.~(1) is not consistent with the detailed electronic structures. 

The results shown in Figs.~1(a)-(b) are representative of the spectral characteristics of both samples, as manifested by the gap spatial maps and the corresponding histograms for both samples in Figs.~2(a)-(b), where the gap values are empirically determined as one-half of the peak-to-peak values. Overall $\langle \Delta _{\Gamma} \rangle = 10.0$ meV and $\langle \Delta _{\rm M} \rangle = 5.0$ meV for $x$ = 0.12 and $\langle \Delta _{\Gamma} \rangle = 8.0$ meV and $\langle \Delta _{\rm M} \rangle = 4.0$ meV for $x$ = 0.06. 
 
\begin{figure}
\centering
\includegraphics[width=3.4in]{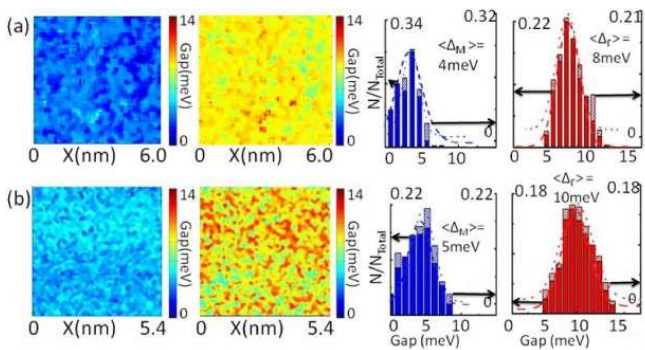}
\caption{(color online) Superconducting gap maps and histograms at T = 6 K: {\bf (a)} Left to right: The first two panels correspond to the $\Delta _{\rm M}$ and $\Delta _{\Gamma}$ maps for the underdoped sample (x = 0.06), and the right two panel represent the corresponding histograms for both the quasiparticle (solid bars) and quasihole (shaded bars) branches, showing particle-hole symmetry and the mean values of $\langle |\Delta _{\rm M}| \rangle = 4$ meV and $\langle |\Delta _{\Gamma}| \rangle = 8$ meV. {\bf (b)} The left two panels are respectively the $\Delta _{\rm M}$ and $\Delta _{\Gamma}$ maps for the sample with x = 0.12. The right two panels are histograms of $\Delta _{\rm M}$ and $\Delta _{\Gamma}$, showing particle-hole symmetry and $\langle |\Delta _{\rm M}| \rangle = 5$ meV, $\langle |\Delta _{\Gamma}| \rangle = 10$ meV.}
\label{fig2}
\end{figure}

Our findings of the two-gap spectra differ from previous STS studies~\cite{BoyerMC08,YinY09}, which may be the result of differences in the surface preparation. Specifically, we show in Fig.~3(a) an example of the atomically resolved surface topography of the overdoped sample over a $\rm (5.4 \times 5.4) \ nm^2$ area. We find that cleaving the samples under argon gas at room temperature generally resulted in fragmented surfaces, as exemplified by the height histogram in Fig.~3(b) for the same $\rm (5.4 \times 5.4) \ nm^2$ area. In particular, we note that the topography in Fig.~3(a) exhibited no apparently reconstructed $(1 \times 2)$ surfaces, and the overall height variations were limited to within one c-axis lattice constant $c_0 = 1.239$ nm. This finding is in contrast to the mostly flat and reconstructed surfaces reported by other groups for cold-cleaved samples~\cite{BoyerMC08,YinY09,ChuangTM10}. Despite the fragmented surfaces, consistent two-gap spectral features were found throughout the scanned area, as exemplified in Fig.~3(c) for the tunneling spectra taken near the middle of Fig.~3(a). 

\begin{figure}
\centering
\includegraphics[width=3.4in]{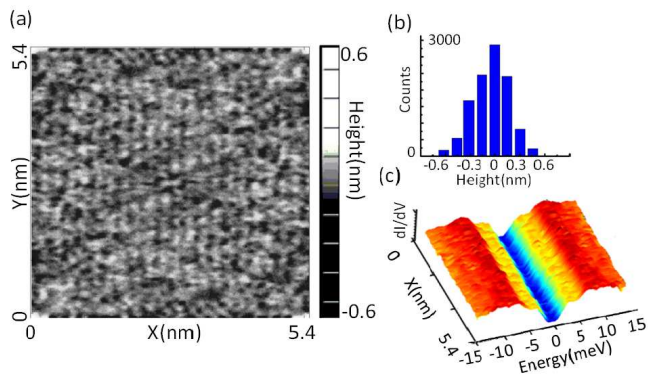}
\caption{(color online) Correlation of atomically resolved surface topography with tunneling conductance spectra: {\bf (a)} Surface topography of the overdoped sample (x = 0.12) over a $\rm (5.4 \times 5.4) nm^2$ area. {\bf (b)} The height histogram of the area in (a), showing height variations within one lattice constant along the c-axis. {\bf (c)} Spatial evolution of the normalized $(dI/dV)$-vs.-$V$ spectra across a horizontal line slightly below the middle of the topography image in (a), which reveals consistent two-gap features.}
\label{fig3}
\end{figure}

Next, we examine the spatial variations in the tunneling conductance at constant bias voltages. In Figs.~4(a)-(b), we illustrate the tunneling conductance maps of two samples and under three different bias voltages of $V$ = 0, $(\langle \Delta _{\rm M} \rangle/e)$ and $(\langle \Delta _{\Gamma} \rangle/e)$. The Fourier transformed (FT) tunneling conductance at $V = (\langle \Delta _{\rm M} \rangle/e)$ and $(\langle \Delta _{\Gamma} \rangle/e)$ is shown in Fig.~4(c) for $x = 0.06$ and in Fig.~4(d) for $x = 0.12$. We find that dominant QPI occur at three wave-vectors~\cite{HanaguriT10}: $q_1$ between two electron pockets across the first Brillouin zone, $q_1 \sim (\pm 2\pi,0)/(0, \pm 2\pi)$; $q_2$ between the hole- and electron-pockets at $\Gamma$ and M-points, $q_2 \sim (\pm \pi,0)/(0, \pm \pi)$; and $q_3$ between two adjacent electron pockets, $q_3 \sim (\pm , \pi, \pm \pi)$. For the underdoped sample, strong QPI at $q_2$ and absence of QPI at $q_3$ is consistent with the five-orbital theoretical calculations~\cite{WangF09a,WangF09b,ZhangYY09} for two possible scenarios: one is scalar-impurity QPI between sign-reversing order parameters associated with the hole- and electron-pockets, and the other is magnetic impurity QPI between order parameters of the same sign. Given that scalar scatterers are generally more common than magnetic impurities, we suggest that the prevailing QPI wave-vectors at $q_2$ are the result of sign-reversing order parameters. In this context, the appearance of $q_3$ wave-vectors in the overdoped sample might be due to a larger density of Co-atoms in the surface Fe/Co layers: Unlike those in the bulk, the charge transfer from the surface Co-atoms is incomplete so that they may behave like magnetic impurities~\cite{WangF09a,WangF09b,ZhangYY09,Plamadeala10}. 

It is worth noting that for both samples the $q_2$ wave-vectors appear to exhibit one preferential direction. In addition, energy-dependent features  at small wave-vectors ($|q| < \pi/2$) are observed along the same direction. These findings may be related to the nematic-order wave-vectors found in the parent state of these compounds~\cite{ChuangTM10}, although our limited momentum resolution cannot provide detailed comparison. Overall, we attribute $q_1$, $q_2$, and $q_3$ to the QPI scattering wave-vectors because they are not only slightly energy dependent but also doping dependent, and so they cannot be simply attributed to Bragg diffractions of the lattices. The relatively weak energy dependence of the QPI wave-vectors is the result of small and nearly isotropic Fermi pockets in the iron arsenides, which differs from the highly energy-dependent QPI wave-vectors in the cuprate superconductors~\cite{McElroy05,Beyer09,Yeh09,Yeh10}. 

\begin{figure}
\centering
\includegraphics[width=3.4in]{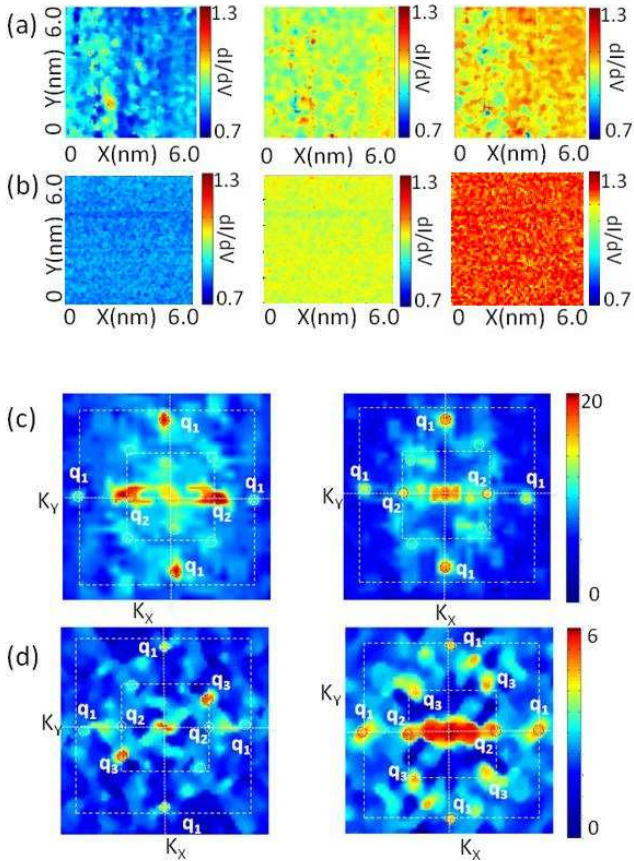}
\caption{(color online) Tunneling conductance maps of $\rm Ba(Fe_{1-x}Co_x)_2As_2$ samples at three constant bias voltages $V = 0$, $(\langle \Delta _{\rm M} \rangle/e)$, $(\langle \Delta _{\Gamma} \rangle /e)$ and for $T$ = 6 K: {\bf (a)} x = 0.06 and {\bf (b)} x = 0.12. {\bf (c)} Fourier transformed (FT) tunneling conductance maps at $V = (\langle \Delta _{\rm M} \rangle /e) = 4$ meV (left panel) and $V = (\langle \Delta _{\Gamma} \rangle /e) = 8$ meV (right panel) for x = 0.06. {\bf (d)} FT conductance maps at $V = (\langle \Delta _{\rm M} \rangle /e) = 5$ meV (left panel) and $V = (\langle \Delta _{\Gamma} \rangle /e) = 10$ meV (right panel) for x = 0.12. The thin white lines define the first and second Brillouin zones, and we have used the lattice constant $a = 0.53$ nm for the length of the first Brillouin zone $(2 \pi /a)$. All three QPI wave-vectors $q_1$, $q_2$ and $q_3$ are slightly energy and doping dependent.}
\label{fig4}
\end{figure}

Finally, we note that the relatively high zero-bias conductance in all tunneling spectra at $T \ll T_c$ is suggestive of dominant unitary impurity scattering~\cite{BangY09}. Further, unitary impurity effects on the suppression of $T_c$ for the sign-changing $s$-wave superconductors are found to be as significant as those on the $d$-wave superconductors~\cite{BangY09}. On the other hand, unitary impurity effects on suppressing the sign-changing $s$-wave order parameters involve sign-dependent components that are partially cancelled, and are therefore weakened~\cite{BangY09}. Hence, the large ratios of $(2 \langle \Delta _{\Gamma , \rm M} \rangle)/(k_B T_c)$ for the 122 system, with $(2 \langle \Delta _{\rm M} \rangle)/(k_B T_c) \sim $ 6.6 (5.8) and $(2 \langle \Delta _{\Gamma} \rangle)/(k_B T_c) \sim$ 13.2 (11.6) for $x$ = 0.06 (0.12), may be attributed to significant unitary impurity scattering in these sign-changing $s$-wave superconductors. A possible source for the unitary impurity scattering may be associated with disorder in the Co doping into the Fe-planes, which is in contrast to the situation in the 1111 iron arsenides where doping takes place in the charge reservior~\cite{RenZA08} so that the disorder effect is weaker and the $(2 \Delta)/(k_B T_c)$ ratio is small ($\sim 3$)~\cite{ChenTY08,FasanoY10}, whereas the $T_c$ values are generally higher.

In summary, we have demonstrated direct STS evidence for two-gap superconductivity in electron-doped $\rm Ba(Fe_{1-x}Co_x)_2As_2$ single crystals of two doping levels. The Fourier transformed tunneling conductance reveals strong quasiparticle scattering interferences near the nesting wave-vector between the hole Fermi pockets at $\Gamma$ and the electron Fermi pockets at M, consistent with sign-changing order parameters of the two Fermi pockets. The excess zero-bias conductance and the large $2\Delta _{\Gamma , \rm M}/(k_B T_c)$ ratios for both doping levels may be attributed to significant unitary impurity scattering in a sign-changing $s$-wave superconducting system.

\begin{acknowledgments}
This work at Caltech was jointly supported by the NSF Grant DMR-0907251, and the Kavli and Moore Foundations. The work in China was supported by the NSFC, the Ministry of Science and Technology of China and Chinese Academy of Sciences within the knowledge innovation program. We thank Patrick A. Lee, Igor Mazin and Yunkyu Bang for valuable discussion.
\end{acknowledgments}


\begin{thebibliography}{27}

\bibitem{Kamihara08}
Y. Kamihara, T. Watanabe, M. Hirano, and H. Hosono, J. Am. Chem. Soc. {\bf 130}, 3296 (2008).

\bibitem{Takahashi08}
H. Takahashi {\it et al.}, Nature {\bf 453}, 376 (2008).

\bibitem{ChenXH08}
X. H. Chen {\it et al.}, Nature {\bf 453}, 761 (2008).

\bibitem{WenHH08}
H.-H. Wen {\it et al.}, Europhys. Lett. {\bf 82}, 17009 (2008).

\bibitem{RenZA08}
Z.-A. Ren {\it et al.}, Europhys. Lett. {\bf 83}, 17002 (2008).

\bibitem{Sasmal08}
K. Sasmal {\it et al.}, Phys. Rev. Lett. {\bf 101}, 107007 (2008).

\bibitem{Rotter08}
M. Rotter {\it et al.}, Phys. Rev. Lett. {\bf 101}, 107006 (2008).

\bibitem{Sefat08}
A. S. Sefat {\it et al.}, Phys. Rev. Lett. {\bf 101}, 117004 (2008).

\bibitem{MasseeF09}
F. Massee {\it et al.}, Phys. Rev. B {\bf 80}, 140507(R) (2009).

\bibitem{KatoT09}
T. Kato {\it et al.}, Phys. Rev. B {\bf 80}, 180507(R) (2009).

\bibitem{Cvetkovic09}
V. Cvetkovic and Z. Tesanovic, Europhys. Lett. {\bf 85}, 37002 (2009).

\bibitem{delaCruz08}
C. de la Cruz {\it et al.}, Nature {\bf 453}, 899 (2008).

\bibitem{SinghDJ08}
D. J. Singh and M.-H. Du, Phys. Rev. Lett. {\bf 100}, 237003 (2008).

\bibitem{Haule08}
K. Haule, J. H. Shim, and G. Kotliar, Phys. Rev. Lett. {\bf 100}, 226402 (2008).

\bibitem{WangF09a}
F. Wang, H. Zhai and D.-H. Lee, Europhys. Lett. {\bf 85}, 37005 (2009).

\bibitem{WangF09b}
F. Wang {\it et al.}, Phys. Rev. Lett. {\bf 102}, 047005 (2009).

\bibitem{MazinII08}
I. I. Mazin, D. J. Singh, M. D. Johannes, and M. H. Du, Phys. Rev. Lett. {\bf 101}, 057003 (2008).

\bibitem{LuDH08}
D. H. Lu {\it et al.}, Nature {\bf 455}, 81 (2008).

\bibitem{DingH08}
H. Ding {\it et al.}, Europhys. Lett. {\bf 83}, 47001 (2008).

\bibitem{TerashimaK09}
K. Terashima {\it et al.}, Proc. Nat. Acad. Sci. {\bf 106}, 7330 (2009).

\bibitem{ChenCT10}
C.-T. Chen {\it et al.}, Nature Phys. {\bf 6}, 260 (2010).

\bibitem{HanaguriT10}
T. Hanaguri {\it et al.}, Science {\bf 328}, 474 (2010).

\bibitem{ChenTY08}
T. Y. Chen {\it et al.}, Nature {\bf 453}, 1224 (2008).

\bibitem{ShanL08}
L. Shan {\it et al.}, Europhys. Lett. {\bf 83}, 57004 (2008).

\bibitem{BoyerMC08}
M. C. Boyer {\it et al.}, arXiv:0806.4400 (2008).

\bibitem{YinY09}
Y. Yin {\it et al.}, Phys. Rev. Lett. {\bf 102}, 097002 (2009).

\bibitem{ChuangTM10}
T.-M. Chuang {\it et al.}, Science {\bf 327}, 181 (2010).

\bibitem{FasanoY10}
Y. Fasano {\it et al.}, Phys. Rev. Lett. {\bf 105}, 167005 (2010).

\bibitem{BangY09}
Y. Bang, H.-Y.Choi and H. Won, Phys. Rev. B {\bf 79}, 054529 (2009).

\bibitem{FangL09}
L. Fang {\it et al.}, Phys. Rev. B {\bf 80}, 140508 (2009).

\bibitem{BernhardC09}
C. Bernhard {\it et al.}, New J. Phys. {\bf 11}, 055050 (2009).

\bibitem{ShenB10}
B. Shen {\it et al.}, Phys. Rev. B {\bf 81}, 014503 (2010).

\bibitem{DynesRC78}
R. C. Dynes, V. Narayanamurti and J. P. Garno, Phys. Rev. Lett. {\bf 41}, 1509 (1978).

\bibitem{ZhangYY09}
Y.-Y. Zhang {\it et al.}, Phys. Rev. B {\bf 80}, 094528 (2009).

\bibitem{Plamadeala10}
E. Plamadeala, T. Pereg-Barnea, G. Refael, Phys. Rev. B {\bf 81}, 134513 (2010).

\bibitem{McElroy05}
K. McElroy {\it et al.}, {\it Phys. Rev. Lett.} {\bf 94}, 197005 (2005).

\bibitem{Beyer09}
A. D. Beyer {\it et al.}, Europhys. Lett. {\bf 87}, 37005 (2009).

\bibitem{Yeh09}
N.-C. Yeh and A. D. Beyer, Int. J. Mod. Phys. B {\bf 23}, 4543 (2009).

\bibitem{Yeh10}
N.-C. Yeh {\it et al.}, J. Supercond. Nov. Magn. {\bf 23}, 757 (2010).

\end{thebibliography}
\end{document}